\documentclass[letter,longauth]{aa} 

\usepackage{graphicx}
\usepackage{txfonts}
\begin{document} 

   \title{Refinement of the convex shape model and tumbling spin state of (99942) Apophis using the 2020-2021 apparition data}

   \titlerunning{Apophis}

   \author{H.-J. Lee\inst{1}\and
    M.-J. Kim\inst{1}\and
    A. Marciniak\inst{2}\and
    D.-H. Kim\inst{1}\fnmsep\inst{3}\and
    H.-K. Moon\inst{1}\and
    Y.-J. Choi\inst{1}\fnmsep\inst{4}\and
    S. Zo{\l}a\inst{5}\and
    J. Chatelain\inst{7}\and
    T. A. Lister\inst{7}\and
    E. Gomez\inst{8}\and
    S. Greenstreet\inst{9}\fnmsep\inst{10}\and
    A. P{\'a}l\inst{11}\and
    R. Szak{\'a}ts\inst{11}\and
    N. Erasmus\inst{12}\and
    R. Lees\inst{13}\and
    P. Janse van Rensburg\inst{12}\fnmsep\inst{13}\and\\
    W. Og{\l}oza\inst{6}\and
    M. Dr{\'o}{\.z}d{\.z}\inst{6}\and
    M. {\.Z}ejmo\inst{14}\and
    K. Kami{\'n}ski\inst{2}\and
    M. K. Kami{\'n}ska\inst{2}\and
    R. Duffard\inst{15}\and
    D.-G. Roh\inst{1}\and
    H.-S. Yim\inst{1}\and
    T. Kim\inst{16}\and
    S. Mottola\inst{17}\and
    F. Yoshida\inst{18}\fnmsep\inst{19}\and
    D. E. Reichart\inst{20}\and
    E. Sonbas\inst{21}\and
    D. B. Caton\inst{22}\and
    M. Kaplan\inst{23}\and
    O. Erece\inst{23}\fnmsep\inst{24}\and
    H. Yang\inst{1}
    }
    
    \institute{Korea Astronomy and Space Science Institute, 776, Daedeokdae-ro, Yuseong-gu, Daejeon 34055, Korea\\ \email{hjlee@kasi.re.kr} \and
    Astronomical Observatory Institute, Faculty of Physics, A. Mickiewicz University, S{\l}oneczna 36, 60-286 Pozna{\'n}, Poland\and
    Chungbuk National University, 1 Chungdae-ro, Seowon-gu, Cheongju, Chungbuk 28644, Korea\and
    University of Science and Technology, 217, Gajeong-ro, Yuseong-gu, Daejeon 34113, Korea\and
    Astronomical Observatory, Jagiellonian University, ul. Orla 171, 30-244 Krak{\'o}w, Poland\and
    Mt. Suhora Observatory, Pedagogical University, ul. Pochor\k{a}{\.z}ych 2,  30-084 Krak{\'o}w, Poland\and
    Las Cumbres Observatory, 6740 Cortona Drive Suite 102, Goleta, CA 93117, USA\and
    Las Cumbres Observatory, School of Physics and Astronomy, Cardiff University, QueensBuildings, The Parade, Cardiff CF24 3AA, UK\and
    Asteroid Institute, 20 Sunnyside Ave, Suite 427, Mill Valley, CA 94941, USA\and
    Department of Astronomy and the DIRAC Institute, University of Washington, 3910 15thAve NE, Seattle, WA 98195, USA\and
    Konkoly Observatory, Research Centre for Astronomy and Earth Sciences, E\H{o}tv\H{o}s Lor{\'a}nd Research Network (ELKH), H-1121 Budapest, Konkoly Thege Mikl{\'o}s {\'u}t 15-17, Hungary\and
    South African Astronomical Observatory, Cape Town, 7925, South Africa\and
    Department of Astronomy, University of Cape Town, Rondebosch, 7701, South Africa\and
    Kepler Institute of Astronomy, University of Zielona G{\'o}ra, Lubuska 2, 65-265 Zielona G{\'o}ra, Poland\and
    Departamento de Sistema Solar, Instituto de Astrof{\'i}sica de Andaluc{\'i}a (CSIC), Glorieta de la Astronom{\'i}a s/n, 18008   Granada, Spain\and
    National Youth Space Center, Goheung, Jeollanam-do 59567, Korea\and
    Deutsches Zentrum f\"ur Luft- und Raumfahrt (DLR), Institute of Planetary Research, D-12489 Berlin, Germany\and
    University of Occupational and Environmental Health, Japan, 1-1 Iseigaoka, Yahata, Kitakyusyu 807-8555, Japan\and
    Planetary Exploration Research Center, Chiba Institute of Technology, 2-17-1 Tsudanuma, Narashino, Chiba, 275-0016, Japan\and
    University of North Carolina at Chapel Hill, Chapel Hill, North Carolina NC 27599, USA\and
    Department of Physics, Adiyaman University, 02040 Adiyaman, Turkey\and
    Dark Sky Observatory, Dept. of Physics and Astronomy, Appalachian State University, Boone, NC 28608, USA\and
    Akdeniz University, Department of Space Sciences and Technologies, 07058 Antalya, Turkey\and
    T\"{U}B\.ITAK National Observatory, Akdeniz University Campus, 07058 Antalya, Turkey
    }
    
   \date{Received ; accepted }
 
  \abstract
   {The close approach of the near-Earth asteroid (99942) Apophis to Earth in 2029 will provide a unique opportunity to examine how the physical properties of the asteroid could be changed due to the Earth’s gravitational perturbation. As a result, the Republic of Korea is planning a rendezvous mission to Apophis.}
   {Our aim was to use photometric data from the apparitions in 2020-2021 to refine the shape model and spin state of Apophis.}
   {Using thirty-six 1 to 2-m class ground-based telescopes and the Transiting Exoplanet Survey Satellite, we performed a photometric observation campaign throughout the 2020-2021 apparition. The convex shape model and spin state were refined using the light-curve inversion method.}
   {According to our best-fit model, Apophis is rotating in a short axis mode with rotation and precession periods of 264.178 hours and 27.38547 hours, respectively. The angular momentum vector orientation of Apophis was found as (275$^\circ$, -85$^\circ$) in the ecliptic coordinate system. The ratio of the dynamic moments of inertia of this asteroid was fitted to $I_{a}:I_{b}:I_{c}=0.64:0.97:1$, which corresponds to an elongated prolate ellipsoid. These findings regarding the spin state and shape model could be used to not only design the space mission scenario but also investigate the impact of the Earth's tidal force during close encounters.}
   {}


   \maketitle
%

\begin{table*}[t]
\caption{Details of the observatories and instruments in this campaign.}
\label{table:1}
\centering
\begin{tabular}{l l l l}     
\hline\hline
Telescope					&	Latitude				&	Longitude				&	Instrument					\\
\hline                                       
\multicolumn{4}{c}{{\bf –Ground-based telescopes-}} \\
Adiyaman Observatory 0.6 m	&	~37:45:06 N				&	~38:13:32 E				&	Andor Tech					\\ 
AMU Winer, RBT 0.7m     	&	~31:39:56 N				&	110:36:06 W				&	Andor iXon					\\ 
ATLAS HKO 0.5 m				&	~20:42:27 N				&	156:15:25 W				&	STA-1600 10.5K CCD			\\ 
ATLAS MLO 0.5 m				&	~19:32:10 N				&	155:34:34 W				&	STA-1600 10.5K CCD			\\ 
BOAO 1.8 m					&	~36:09:53 N				&	128:58:36 E				&	E2V 4K CCD					\\ 
CAHA 1.23 m					&	~37:13:25 N				&	~~2:32:46 W				&	DLR MKIII camera			\\ 
CAHA 2.2 m					&	~37:13:25 N				&	~~2:32:46 W				&	CAFOS						\\ 
DOAO 1.0 m                  &   ~34:31:35 N             &   127:26:48 E             &   PI SOPHIA-2048B CCD         \\ 
Kawabe Cosmic Park 1.0 m	&	~33:53:27 N				&	135:13:12 E				&	FLI PL09000					\\ 
KMTNet CTIO 1.6 m			&	~30:10:02 S				&	~70:48:14 W				&	18K mosaic CCD with four E2V 9K				\\ 
KMTNet SAAO 1.6 m			&	~32:22:43 S				&	~20:48:37 E				&	18K mosaic CCD with four E2V 9K 				\\ 
KMTNet SSO 1.6 m			&	~31:16:16 S				&	149:03:45 E				&	18K mosaic CCD with four E2V 9K				\\ 
Krakow-CDK500 0.5 m			&	~50:03:15 N				&	~19:49:41 E				&	Apogee USB/Net				\\ 
LCO CTIO A 1.0 m			&	~30:10:03 S				&	~70:48:17 W				&	Sinistro					\\ 
LCO CTIO B 1.0 m			&	~30:10:03 S				&	~70:48:17 W				&	Sinistro					\\ 
LCO McDonald A 1.0 m		&	~30:41:47 S				&	104:00:54 E				&	Sinistro					\\ 
LCO McDonald B 1.0 m		&	~30:41:48 S				&	104:00:54 E				&	Sinistro					\\ 
LCO SAAO A 1.0 m			&	~32:23:50 N 			&	~20:48:37 W				&	Sinistro					\\ 
LCO SAAO B 1.0 m			&	~32:23:50 N 			&	~20:48:36 W				&	Sinistro					\\ 
LCO SAAO C 1.0 m			&	~32:23:51 N 			&	~20:48:36 W				&	Sinistro					\\ 
LCO SSO A 1.0 m				&	~31:16:22 N				&	149:04:14 W				&	Sinistro					\\ 
LCO SSO B 1.0 m				&	~31:16:23 N				&	149:04:15 W				&	Sinistro					\\ 
LOAO 1.0 m					&	~32:26:32 N				&	110:47:19 E				&	E2V 4K CCD					\\ 
OWL Mitzpeh Ramon 0.5 m		&   ~30:35:51 N				&	~34:45:48 E				&	FLI 16803					\\ 
OWL Oukaimeden 0.5 m		&   ~31:12:21 N				&	~~7:52:00 W				&	FLI 16803					\\ 
OWL Tucson 0.5 m			&   ~32:26:31 N				&	110:47:21 W				&	FLI 16803					\\ 
OWL Yeongcheon 0.5 m		&   ~36:09:50 N				&	128:58:33 E				&	FLI 16803					\\ 
SAAO Lesedi 1.0 m			&	~32:22:47 S				&	~20:48:36 E				&	SHOC (Andor iXon 888)		\\ 
Skynet DSO-14, 0.4 m		&	~36:15:01 N				&	~81:24:45 W				&	Apogee USB/Net				\\ 
Skynet Prompt5, 0.4 m		&	~30:10:03 S				&	~70:48:19 W				&	Apogee USB/Net				\\ 
Skynet Prompt6, 0.4 m		&	~30:10:03 S				&	~70:48:19 W				&	FLI							\\ 
Skynet Prompt MO 1 0.4 m	&	~31:38:18 S				&	116:59:19 E				&	Apogee USB/Net				\\ 
Skynet RRRT 0.6 m			&	~37:52:44 N				&	~78:41:39 W				&	SBIG STX-16803 3 CCD		\\ 
SOAO 0.6 m					&	~36:56:04 N				&	128:27:27 E				&	FLI 4K						\\ 
Suhora Observatory Zeiss-60	&	~49:34:09 N				&	~20:04:03 E				&	Apogee AltaU-47 			\\ 
TUG 1.0 m					&	~36:49:27 N				&	~30:20:08 E				&	SI 4K						\\ 
\multicolumn{4}{c}{{\bf –Space-based telescope-}} \\
TESS 0.1 m              	&	        				&	        				&	four MIT Lincoln Lab. CCID-80 devices          \\ 
\hline
\end{tabular}
\tablefoot{AMU = Adam Mickiewicz University, RBT = Roman Baranowski Telescope,
           ATLAS = Asteroid Terrestrial-impact Last Alert System, HKO = Haleakala Observatory, MLO = Mauna Loa Observatory, 
           BOAO = Bohyunsan Optical Astronomy Observatory, DOAO = Deokheung Optical Astronomy Observatory,
		   CAHA = Calar Alto Observatory, KMTNet = Korea Microlensing Telescope \citep{Kim_et_al_2016}, CTIO = Cerro Tololo Inter-American Observatory, 
		   SAAO = South African Astronomical Observatory, SSO = Siding Spring Observatory, LCO = Las Cumbres Observatory, LOAO = Lemonsan Optical Astronomy Observatory, 
		   CDK = Corrected Dall-Kirkham, OWL = Optical Wide-field patroL Network, DSO = Dark Sky Observatory, MO = Meckering Observatory, 
		   RRRT = Rapid Response Robotic Telescope, SOAO = Sobaeksan Optical Astronomy Observatory, TUG = T\"{U}B\'{I}TAK National observatory, TESS = Transiting Exoplanet Survey Satellite, CAFOS = Calar Alto Faint Object Spectrograph}
\end{table*}

\section{Introduction} \label{sec:intro}

The Aten-type near-Earth asteroid (99942)~Apophis (2004~MN4) (hereafter Apophis) is an Sq-type asteroid with an estimated size of 340~m \citep{Binzel_et_al_2009,Brozovic_et_al_2018,Reddy_et_al_2018}. This asteroid was discovered on June 19, 2004, by R.A. Tucker, D.J. Tholen, and F. Bernardi at Kitt Peak, Arizona. From early prediction, it was estimated that this asteroid could impact Earth with a maximum probability of 2.7\% (JPL Sentry on December 27, 2004; \citealt{Chesley_2006}). As the accuracy of the orbital prediction improved in follow-up observations, the impact probability of Apophis was reduced. In particular, the prediction derived from high-precision radar observations at the Arecibo Observatory in 2005 and 2006 indicated that this asteroid will pass by Earth at a geocentric distance of 38,326 km (approximately 6 Earth radii), which is within the geosynchronous orbit in April 2029 \citep{Giorgini_et_al_2008}. Recently, the radar observations at the Goldstone Solar System Radar (GSSR) and Green Bank Telescope (GBT) in March 2021 have precisely estimated Apophis' orbit around the Sun, ruling out any Earth impact threat for the next hundred years or more  \citep{Greicius}.

Although Apophis' impact threat has disappeared, this asteroid remains an object of interest because of its close approach in 2029. The Earth encounter of Apophis in 2029 is expected to trigger varying degrees of alterations in the dynamics, spin-states, and surface arrangements of Apophis due to the Earth's gravitational perturbation \citep{Yu_et_al_2014, Souchay_et_al_2014,Souchay_et_al_2018, DeMartini_et_al_2019, Hirabayashi_et_al_2021, Valvano_et_al_2022}. Thus, the study of this asteroid will provide an excellent opportunity to examine the evolutionary process of its physical properties caused by planetary perturbation. For this reason, Apophis became a unique observation target and is the primary mission target of the Rendezvous Mission to Apophis, which is currently under pre-Phase A study in the Republic of Korea and scheduled for launch in 2027 \citep{Moon_et_al_2020}.

The shape model and spin state are the most fundamental parameters for predicting the evolutionary process due to Earth's tidal effect. In addition, these properties provide important information for planning space mission scenarios. The spin state and convex shape model of Apophis were reconstructed using photometric data obtained from the 2012-2013 apparition \citep{Pravec_et_al_2014}. They found it has non-principal axis (NPA) rotation in a short-axis mode (SAM) with rotation and precession periods of 263 and 27.38 hours, respectively, and the orientation of angular momentum vector of $\lambda_{L}$ = 250$^\circ$ and $\beta_{L}$ = -75$^\circ$. In addition, the convex shape model of \cite{Pravec_et_al_2014} can be approximated by a prolate ellipsoid with a ratio of the greatest and intermediate principal moments of inertia ($I_{b}/I_{c}$) of 0.965, and the smallest principal moment of inertia ($I_{a}/I_{c}$) of 0.61.

For the last time before the 2029 Earth encounter, Apophis approached Earth on March 2021 at 0.11 AU and its apparent brightness increased to reach magnitude 16. Thus, the observation window for Apophis from the end of 2020 to the beginning of 2021 was the last opportunity to investigate its spin properties and refine the convex shape model. Therefore, we planned a photometric observation campaign for Apophis during this apparition. The details of our observation campaign are described in Section~\ref{sec:obs}. In Section~\ref{sec:result}, a period analysis and reconstruction of the spin state and shape model of Apophis are reported. Finally, the summary and conclusions of this letter are given in Section~\ref{sec:Summary_and_Conclusions}.

   \begin{figure*}[t]
   \centering
    \includegraphics[width=140mm]{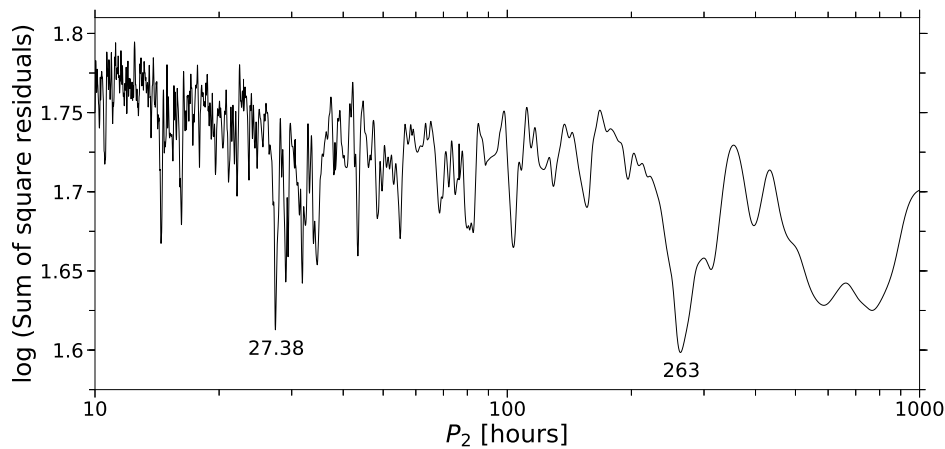}
      \caption{$P_{2}$ search diagram of Apophis. The sum of square residuals were calculated for the 4-th order two-period Fourier series with $P_{1}$ = 30.56 hours fitted to our dense light curve obtained from 2021-02-07.0 to 2021-03-16.2 in flux units.}
         \label{fig:P2_search}
   \end{figure*}

   \begin{figure*}[t]
   \centering
    \includegraphics[width=146mm]{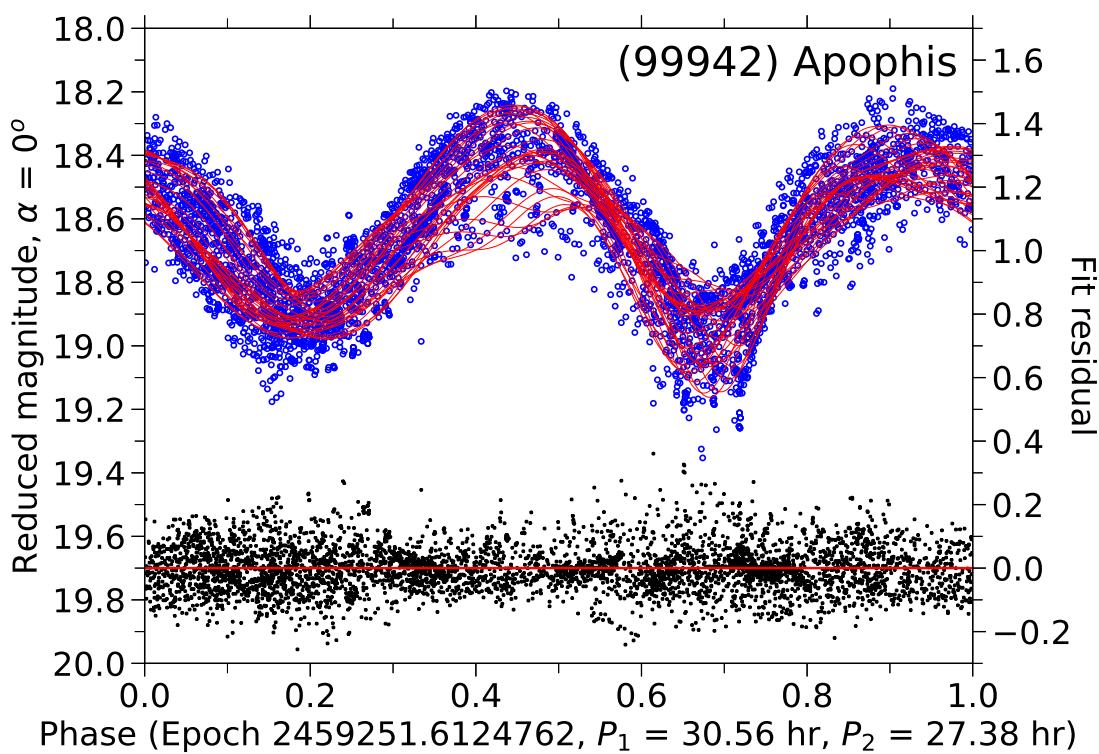}
      \caption{The light curve of Apophis taken from 2021-02-07.0 to 2021-03-16.2 reduced to the unit geo- and heliocentric distance and to a consistent solar phase angle. The blue open circles indicate the photometric data folded with $P_1$. The red curves denote the best-fit 4th order two-period Fourier series with the periods $P_{1}$ = 30.56 hours and $P_{2}$ = 27.38 hours. The black squares indicate the residuals of the photometric data from the 4th order two-period Fourier series (see the right ordinates).}
         \label{fig:lc}
   \end{figure*}
   \begin{figure*}[t]
   \centering
   \includegraphics[width=1.1\textwidth, trim=7cm 5cm 0cm 3cm, clip]{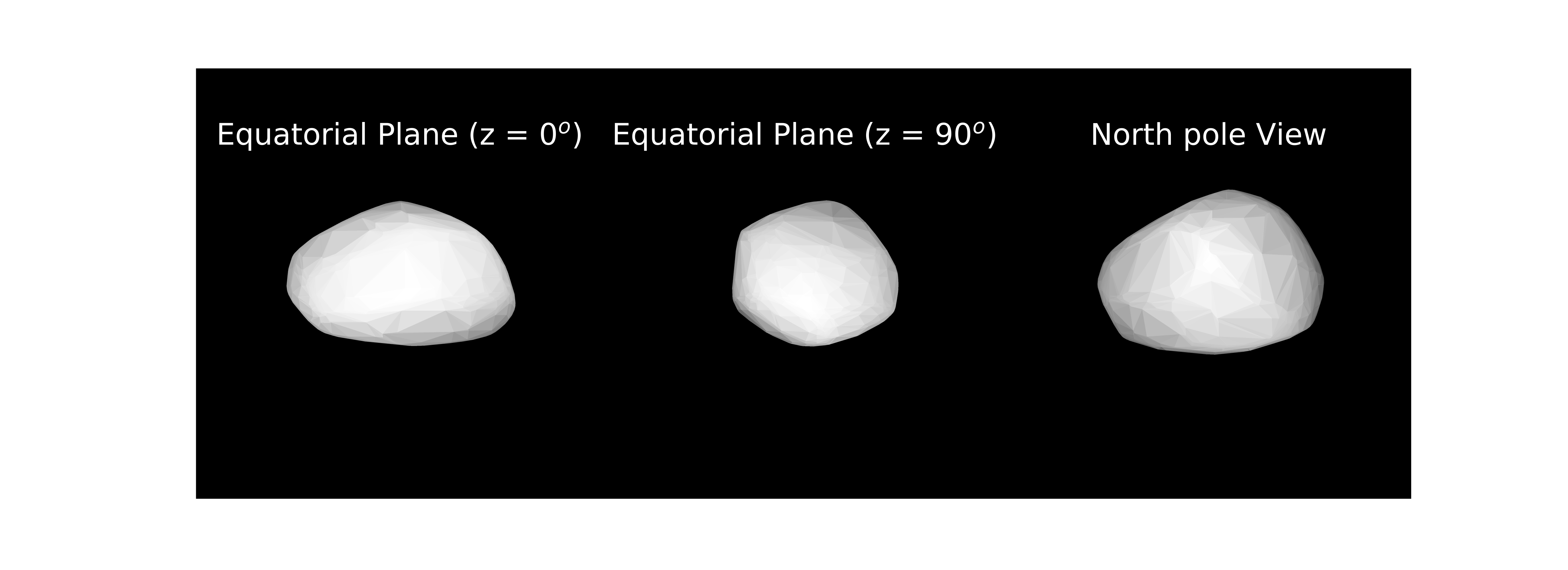}                                          
      \caption{Convex shape model of Apophis.}
        \label{fig:shape}
   \end{figure*}

\section{Photometric observation campaign in the 2020-2021 apparition} \label{sec:obs}

As mentioned above, the 2020-2021 apparition of Apophis was the last opportunity to reveal the physical properties of this asteroid before its 2029 close approach. Therefore, we organized an extensive and long-term photometric observation campaign for Apophis during this apparition. Our observation campaign was conducted by using not only the ground-based telescope but also the space telescope that is Transiting Exoplanet Survey Satellite \citep[TESS;][]{Ricker_et_al_2015} spacecraft. The details of telescopes and instruments used in our campaign are provided in Table~\ref{table:1}. The geometries and observational circumstances are listed in Table~\ref{table:A1}. This campaign’s data also contributed to the global Apophis Planetary Defense Campaign \citep{Reddy_et_al_2022_submitted}.

Ground-based observations were carried out in 11 countries, including the Republic of Korea, the US, Chile, South Africa, Australia, Poland, Spain, Turkey, and Japan using 36 telescopes. Through our observation campaign, we observed 214 dense-in-time light curves and two sparse-in-time light curves. The dense-in-time light curves were obtained using a Johnson-Cousins $V$ or $R$ filter except for the data from the Kawabe Cosmic Park, which used the SDSS $r'$ filter. As the observations were made with different telescopes and instruments, the data reduction processes may be slightly different. All data reductions were conducted following the standard manner. The bias- and flat-field images corrected, and the instrument magnitudes for each frame were measured using aperture photometry. All photometric data were calibrated with ATLAS Refcat2 \citep[The ATLAS All-Sky Stellar Reference Catalog;][]{Tonry_et_al_2018}. The magnitudes of ATLAS Refcat2 were converted to Johnson–Cousins $V$ and $R$ magnitudes using empirical transformation equations \citep{Tonry_et_al_2012}. On the other hand, the sparse-in-time light curves were obtained from ATLAS \citep{Tonry_et_al_2018,Tonry_2018_PASP}. The ATLAS light curve was observed between November 2020 and April 2021 using orange ($o$, 560–820 nm) and cyan ($c$, 420-650 nm) filters. 

We also gathered the long-term continuous photometric data observed from the TESS spacecraft using a wide $I$ passband (see Figure 1 in \citet{Ricker_et_al_2015}). TESS photometric data have been obtained in a similar manner as it was done in \citep{Pal_et_al_2020} for the image series of Sector 35 as acquired between 2021-02-19 and 2021-03-07. The individual photometric data points were derived using a convolution-based differential image analysis by employing the tools of the FITSH package \cite{Pal_2012}. 

\section{Results} \label{sec:result}

\subsection{Periodic analysis}

Before the shape model and spin state of Apophis were reconstructed, we performed periodic analysis of the light curve obtained from our observation campaign. To minimize the possible systematic effects caused by the changes in the observing geometry, this analysis was only conducted using the dense-in-time light curves observed with phase angles from 20$^\circ$ to 40$^\circ$, which corresponds to the period from 2021-02-07.0 to 2021-03-16.2. Because our observations were made with different filters, we corrected them to match the R-filter using the color indices $V - R$ = 0.38, $R - T$ = 0.07, and $r' - R$  = 0.33. The data were converted to flux units, and the helio- and geocentric distances were corrected to the unit distance and the solar phase angle to a consistent value using the H-G phase relation assuming G = 0.24. As Apophis exhibited tumbling motion, we attempted to detect double periods from the light curve of this asteroid.

First, the Lomb-Scargle method \citep{Lomb_1976, Scargle_1982} was adopted to search for periodicities in the light curve. The strongest signal in the Lomb-Scargle periodogram was found at a period of 15.28 hours. The Lomb-Scargle method performs periodic analysis based on a single peak light curve. However, most light curves of asteroids have double peaks because their shapes are elongated. Therefore, we determined the primary period ($P_{1}$) on the light curve of Apophis as 30.56 hours that is double the strongest signal in the Lomb-Scargle periodogram.

The secondary period ($P_{2}$) was determined using the two-period Fourier series method \citep{Pravec_et_al_2005, Pravec_et_al_2014}. The two-period Fourier series employed in this analysis are presented as follows:

\begin{align*}
&F(t) = C_{0} + \sum\limits_{j=1}^{m}\Bigg[ C_{j0} \cos{\frac{2 \pi j}{P_{1}}}t + S_{j0} \sin{\frac{2 \pi j}{P_{1}}}t \Bigg] \\&\qquad\; + \sum\limits_{k=1}^{m}\sum\limits_{j=-m}^{m}\Bigg[C_{jk} \cos{(\frac{2 \pi j}{P_{1}} + \frac{2 \pi k}{P_{2}})}t + S_{jk} \sin{(\frac{2 \pi j}{P_{1}} + \frac{2 \pi k}{P_{2}})}t\Bigg]
\end{align*}

The result for the $P_{2}$ search is shown in Figure~\ref{fig:P2_search}, where the abscissa is $P_{2}$ and the ordinate is the sum of the squared residuals for the fitted 4-th order two-period Fourier series with $P_{1}$ = 30.56 hours. From our $P_{2}$ search, we found two minimums at 27.38 and 263 hours. As the long period of 263 hours was a combination of the short period of 27.38 hours with $P_{1}$, that is, $263^{-1} \approx 27.38^{-1} - 30.56^{-1}$, it seems to be derived from the short period. Thus, we determined $P_{2}$ as 27.38 hours. Figure~\ref{fig:lc} shows a composite light curve of Apophis obtained from 2021-02-07.0 to 2021-03-16.2 with the fitted 4-th order two-period Fourier series for periods 30.56 and 27.38 hours.
 
\subsection{Reconstruction of the convex shape model and spin state from light curves} \label{sec:PM}

   \begin{table}[t]
      \caption{Parameters of the Apophis model.}
         \label{table:2}
            \centering
            \begin{tabular}{l l l}     
            \hline\hline
                            	                                 & This Work			    & \cite{Pravec_et_al_2014}		\\
            \hline
            $\lambda_{L}$ [deg]                                  & $278^{+9}_{-8}$		    &	$250 \pm 27$				\\
            $\beta_{L}$ [deg]                                    & $-86^{+5}_{-4}$		&	$-75 \pm 14$          		\\
            $P_{\psi}$ [hour]                                    & $264.18 \pm 0.03$  	&	$263 \pm 6$					\\
            $P_{\phi}$ [hour]                                    & $27.3855 \pm 0.0003$   &	$27.38 \pm 0.07$			\\
            $\psi_{0}$$^{a}$ [deg]                               & $3^{+5}_{-1}$		        &	$14^{+44}_{-11}$			\\
            $\phi_{0}$$^{a}$ [deg]                               & $183^{+7}_{-4}$			    &	$152^{+173}_{-64}$			\\
            $I_{a}/I_{c}$                                        & $0.64^{+0.02}_{-0.09}$	           	&	$0.61^{+0.11}_{-0.08}$		\\
            $I_{b}/I_{c}$                                        & $0.962^{+0.023}_{-0.002}$           	&	$0.965^{+0.009}_{-0.015}$	\\
            $a_\mathrm{dyn}/c_\mathrm{dyn}$                      & $1.48 \pm 0.19$               &    1.51 $\pm$ 0.18          \\
            $b_\mathrm{dyn}/c_\mathrm{dyn}$                      & $1.06 \pm 0.04$               &    1.06 $\pm$ 0.02          \\
            $a_\mathrm{shape}/c_\mathrm{shape}$                  & $1.56 \pm 0.04$          &    1.64 $\pm$ 0.09          \\
            $b_\mathrm{shape}/c_\mathrm{shape}$                  & $1.12 \pm 0.03$          &    $1.14^{+0.04}_{-0.08}$    \\
            $E/E_0$                                              & $1.018 \pm 0.010$        &    $1.024 \pm 0.013$		\\
            \hline
            \end{tabular}
            \tablefoot{$\lambda_{L}$ and $\beta_{L}$ are the ecliptic coordinates of the constant angular momentum vector $\mathbf{L}$; $\psi_{0}$ and $\phi_{0}$ are the standard Euler angles at epoch JD$_{0}$; $I_{a}$, $I_{b}$ and $I_{c}$ are the dynamical moments of inertia corresponding to the longest-, intermediate-, and shortest-axes; $a_\mathrm{dyn}/c_\mathrm{dyn}$ and $c_\mathrm{dyn}/c_\mathrm{dyn}$ are the axial ratios of a dynamically equivalent ellipsoid; $a_\mathrm{shape}/c_\mathrm{shape}$ and $c_\mathrm{shape}/c_\mathrm{shape}$ are the axial ratios of a convex shape model; $E/E_0$ is the ratio of the rotational kinetic energy to the lowest energy for the given angular momentum.\\
            $^a$ The epoch JD$_{0}$ is 2456284.676388 (2012-12-23.176388).}
\end{table}

Given that the radar observation of Apophis suggested that it has a bifurcated shape \citep{Brozovic_et_al_2018}, the non-convex shape model may be a more suitable description of its actual shape. Nonetheless, this non-convex model obtained using photometric data should be applied only after very careful consideration. It is generally not possible to uniquely reconstruct a non-convex model using only photometric data \citep{Viikinkoski_et_al_2017, Harris_Warner_2020}. Further, concavities can be revealed only when reconstruction is performed using data observed at sufficiently high phase angles \citep{Durech_Kaasalainen_2003}. Unfortunately, we did not obtain sufficient data at high phase angles to create a non-convex model of Apophis. Therefore, our analysis was conducted based on the convex shape model.

The convex shape model and spin state of Apophis were reconstructed using the light curve inversion method for the NPA rotator \citep{Kaasalainen_Torppa_2001, Kaasalainen_et_al_2001, Kaasalainen_2001} combined with Hapke's light-scattering model \citep{Hapke_1993}. In this work, we used not only the light curves obtained through our campaign observation but also all available observation data collected from the literature. The historical light curves are listed in Table~\ref{table:A2}.

The first step in revealing the spin state of a tumbling asteroid is to determine the physical periods, that is, the rotation ($P_{\psi}$) and precession period ($P_{\phi}$). Because the periods on the light curve were derived from the physical periods, $P_{1}^{-1}$ and $P_{2}^{-1}$ usually appear at $P_{\phi}^{-1}$ and $P_{\phi}^{-1} \pm P_{\psi}^{-1}$, where the plus sign is for the long-axis mode (LAM) and the minus sign for SAM \citep{Kaasalainen_2001}. Therefore, we found two possible physical period combinations using the period obtained in the previous section: $P_{1}^{-1}  = P_{\phi}^{-1}$ and $P_{2}^{-1} = P_{\phi}^{-1} + P_{\psi}^{-1}$ (LAM); $P_{1}^{-1} = P_{\phi}^{-1} - P_{\psi}^{-1}$ and $P_{2}^{-1} = P_{\phi}^{-1}$ (SAM). The optimization for each $P_{\phi}$ was performed in the same way as in \cite{Lee_et_al_2020, Lee_et_al_2021}. In these produces, the Hapke's model parameters were set to a typical S-type asteroid's: $\varpi = 0.23$, $g = -0.27$, $h = 0.08$, $B_{0} = 1.6$, and $\bar{\theta} = 20^\circ$ \citep{Li_et_al_2015}. Because we did not obtain the data observed at low solar phase angles, the parameters for the opposition surge, $h$ and $B_{0}$, and the roughness $\bar{\theta}$ were fixed. Only the $\varpi$ and $g$ parameters were optimized. It was found that the inertia tensor of the convex shape model for the LAM solution was not consistent with the kinematic $I_1$ and $I_2$ parameters. Therefore, we decided to use SAM as the final solution.

The best-fit model for Apophis is listed in Table~\ref{table:2}, together with \cite{Pravec_et_al_2014} solution for comparison. Further, the uncertainties of our model parameters corresponded to a 3~$\sigma$ confidence interval. The confidence interval was estimated from the increase in the $\chi^2$ value when the solved-for physical parameters were varied. The threshold corresponding to the 3~$\sigma$ confidence interval was set as $\chi^{2}_{min} \times (1 + 3 \sqrt{2/\nu})$\footnote{These was a typo in \cite{Vokrouhlicky_et_al_2017}, so we used $2/\nu$ instead of $2 \nu$.}, where the $\chi^{2}_{min}$ represents the $\chi^{2}$ value for the best-fit solution and $\nu$ represents the number of degrees of freedom \citep{Vokrouhlicky_et_al_2017}. The convex shape model of Apophis is shown in Figure~\ref{fig:shape}. The synthetic light curve of our model and the observed light curve are presented in Appendix~\ref{app:lcs}.

\section{Summary and Conclusions} \label{sec:Summary_and_Conclusions}

In this letter, we present the convex shape model and spin state of Apophis reconstructed using historical and newly obtained light curves. The new photometric data were observed from an extensive and long-term photometric observation campaign for Apophis during its 2020-2021 apparition using 36 facilities, including the ground-based and space telescope. We obtained 211 dense-in-time light curves and two spare-in-time light curves. In the period analysis conducted with our dense light curve, double periods were detected as 30.56 and 27.38 hours, respectively.

The best-fit solution indicates that Apophis is in a SAM state with rotation and precession periods of 264.178 $\pm$ 0.01 and 27.38547 $\pm$ 0.00002 hours, respectively. The ecliptic coordinates of the angular momentum vector orientation of this asteroid are (275$\pm$3$^\circ$, -85$\pm$1$^\circ$). In addition, the ratio of the dynamical moments of inertia was estimated as $I_{a}/I_{c}$ = 0.64 and $I_{b}/I_{c}$ = 0.96. 
Our model was similar to that of \cite{Pravec_et_al_2014}. Nonetheless, the uncertainties of the model parameters were improved because they were reconstructed based on the dataset obtained from the two apparitions. This model will be useful not only for investigating changes in Apophis' physical properties due to the tidal effect during its encounter in 2029 but also for planning a space mission to this asteroid.

\begin{acknowledgements}
This research has made use of the KMTNet system operated by the Korea Astronomy and Space Science Institute (KASI) and the data were obtained at three host sites of CTIO in Chile, SAAO in South Africa, and SSO in Australia. This paper was partially based on observations obtained at the Bohyunsan Optical Astronomy Observatory(BOAO), the Sobaeksan Optical Astronomy Observatory(SOAO), and the Lemmonsan Optical Astronomy Observatory(LOAO), which is operated by the Korea Astronomy and Space Science Institute (KASI). This work has made use of data from the Asteroid Terrestrial-impact Last Alert System (ATLAS) project. The Asteroid Terrestrial-impact Last Alert System (ATLAS) project is primarily funded to search for near earth asteroids through NASA grants NN12AR55G, 80NSSC18K0284, and 80NSSC18K1575; byproducts of the NEO search include images and catalogs from the survey area. The ATLAS science products have been made possible through the contributions of the University of Hawaii Institute for Astronomy, the Queen’s University Belfast, the Space Telescope Science Institute, the South African Astronomical Observatory, and The Millennium Institute of Astrophysics (MAS), Chile. A.P. and R.S. were supported by the K-138962 grant of the National Research, Development and Innovation Office. R.D. acknowledge financial support from the State Agency for Research of the Spanish MCIU through the “Center of Excellence Severo Ochoa” award for the Instituto de Astrofísica de Andalucía (SEV-2017-0709). Based on observations collected at Centro Astronómico Hispano en Andalucía (CAHA) at Calar Alto, operated jointly by Instituto de Astrofísica de Andalucía (CSIC) and Junta de Andalucía.

\end{acknowledgements}

\bibliographystyle{aa}
\bibliography{bibliograpahy_all}

\onecolumn
\begin{appendix} 

\section{Additional tables}
\centering
\begin{longtable}{llllllll}
\caption{The geometries and observational circumstances.} 
\label{table:A1}\\
\hline\hline
            Date UT   & R. A. & Dec.  & $r_{h}$    &  $\Delta$ & $\alpha$    & Observatory & Filter\\
                      & (h m) & ($^\circ$ ')  & (AU)  & (AU)  & ($^\circ$)  &  &      \\
\hline
\endfirsthead
\caption{continued.}\\
\hline\hline
            Date UT   & R. A. & Dec.  & $r_{h}$    &  $\Delta$ & $\alpha$    & Observatory & Filter\\
                      & (h m) & ($^\circ$ ')  & (AU)  & (AU)  & ($^\circ$)  &  &     \\
\hline
\endhead
\hline
\endfoot
\multicolumn{7}{c}{{\bf -–Dense photometry–- }} \\
2021-01-16.5	&	11 44	&	-17 26	&	1.076	&	0.203	&	58.2	&	LOAO 					&	$R$ \\
2021-01-17.5	&	11 43	&	-17 37	&	1.077	&	0.200	&	57.5	&	LOAO 					&	$R$ \\
2021-01-18.5	&	11 43	&	-17 48	&	1.078	&	0.198	&	56.7	&	LOAO 					&	$R$ \\
2021-01-29.1	&	11 34	&	-19 16	&	1.091	&	0.170	&	48.1	&	SAAO Lesedi 			&	$VR$ \\
2021-02-05.0	&	11 21	&	-19 33	&	1.096	&	0.153	&	41.4	&	SAAO Lesedi 			&	$V$ \\
2021-02-05.4	&	11 20	&	-19 33	&	1.096	&	0.152	&	41.0	&	Skynet Prompt5 			&	$R$ \\
2021-02-06.0	&	11 19	&	-19 31	&	1.096	&	0.151	&	40.3	&	SAAO Lesedi 			&	$V$ \\
2021-02-06.3	&	11 18	&	-19 31	&	1.096	&	0.150	&	40.0	&	OWL USA 				&	$R$ \\
2021-02-07.0	&	11 16	&	-19 29	&	1.097	&	0.148	&	39.3	&	CAHA 2.2 m 				&	$R$ \\
2021-02-07.1	&	11 16	&	-19 28	&	1.097	&	0.148	&	39.2	&	LCO SAAO A 				&	$R$ \\
2021-02-07.1	&	11 16	&	-19 28	&	1.097	&	0.148	&	39.2	&	LCO SAAO B 				&	$R$ \\
2021-02-07.1	&	11 16	&	-19 28	&	1.097	&	0.148	&	39.2	&	SAAO Lesedi 			&	$V$ \\
2021-02-07.2	&	11 16	&	-19 28	&	1.097	&	0.148	&	39.1	&	LCO CTIO A 				&	$R$ \\
2021-02-07.3	&	11 15	&	-19 28	&	1.097	&	0.148	&	38.9	&	OWL USA 				&	$R$ \\
2021-02-07.3	&	11 15	&	-19 28	&	1.097	&	0.148	&	38.9	&	Skynet Prompt5 			&	$R$ \\
2021-02-07.8	&	11 14	&	-19 26	&	1.097	&	0.147	&	38.4	&	LCO SAAO C 				&	$R$ \\
2021-02-08.0	&	11 14	&	-19 25	&	1.097	&	0.146	&	38.2	&	LCO SAAO A 				&	$R$ \\
2021-02-08.0	&	11 14	&	-19 25	&	1.097	&	0.146	&	38.2	&	LCO SAAO B 				&	$R$ \\
2021-02-08.1	&	11 13	&	-19 24	&	1.097	&	0.146	&	38.1	&	SAAO Lesedi 			&	$V$ \\
2021-02-08.3	&	11 13	&	-19 23	&	1.097	&	0.145	&	37.8	&	LCO CTIO A 				&	$R$ \\
2021-02-08.3	&	11 13	&	-19 23	&	1.097	&	0.145	&	37.8	&	LCO McDonald A 			&	$R$ \\
2021-02-08.3	&	11 13	&	-19 23	&	1.097	&	0.145	&	37.8	&	OWL USA 				&	$R$ \\
2021-02-08.3	&	11 13	&	-19 23	&	1.097	&	0.145	&	37.8	&	Skynet Prompt5 			&	$R$ \\
2021-02-08.6	&	11 12	&	-19 22	&	1.097	&	0.145	&	37.5	&	LCO SSO B 				&	$R$ \\
2021-02-08.9	&	11 11	&	-19 20	&	1.098	&	0.144	&	37.2	&	LCO SAAO A 				&	$R$ \\
2021-02-09.1	&	11 10	&	-19 19	&	1.098	&	0.144	&	37.0	&	LCO SAAO B 				&	$R$ \\
2021-02-09.1	&	11 10	&	-19 19	&	1.098	&	0.144	&	37.0	&	SAAO Lesedi 			&	$V$ \\
2021-02-09.3	&	11 10	&	-19 17	&	1.098	&	0.143	&	36.7	&	LCO McDonald B 			&	$R$ \\
2021-02-09.3	&	11 10	&	-19 17	&	1.098	&	0.143	&	36.7	&	Skynet Prompt5 			&	$R$ \\
2021-02-09.4	&	11 10	&	-19 17	&	1.098	&	0.143	&	36.6	&	OWL USA 				&	$R$ \\
2021-02-09.6	&	11 09	&	-19 15	&	1.098	&	0.142	&	36.4	&	LCO SSO B 				&	$R$ \\
2021-02-09.9	&	11 08	&	-19 13	&	1.098	&	0.142	&	36.1	&	LCO SAAO C 				&	$R$ \\
2021-02-10.0	&	11 08	&	-19 12	&	1.098	&	0.142	&	35.9	&	LCO SAAO A 				&	$R$ \\
2021-02-10.1	&	11 07	&	-19 12	&	1.098	&	0.141	&	35.8	&	LCO SAAO B 				&	$R$ \\
2021-02-10.1	&	11 07	&	-19 12	&	1.098	&	0.141	&	35.8	&	OWL MAR 				&	$R$ \\
2021-02-10.1	&	11 07	&	-19 12	&	1.098	&	0.141	&	35.8	&	SAAO Lesedi 			&	$V$ \\
2021-02-10.2	&	11 07	&	-19 11	&	1.098	&	0.141	&	35.7	&	Skynet Prompt5 			&	$R$ \\
2021-02-10.3	&	11 07	&	-19 10	&	1.098	&	0.141	&	35.6	&	LCO CTIO B 				&	$R$ \\
2021-02-10.4	&	11 07	&	-19 09	&	1.098	&	0.141	&	35.5	&	LCO McDonald B 			&	$R$ \\
2021-02-10.5	&	11 06	&	-19 08	&	1.098	&	0.140	&	35.4	&	LCO SSO B 				&	$R$ \\
2021-02-11.0	&	11 05	&	-19 04	&	1.098	&	0.139	&	34.8	&	Adiyaman Observatory 	&	$R$ \\
2021-02-11.0	&	11 05	&	-19 04	&	1.098	&	0.139	&	34.8	&	LCO SAAO C 				&	$R$ \\
2021-02-11.0	&	11 05	&	-19 04	&	1.098	&	0.139	&	34.8	&	OWL ISR 				&	$R$ \\
2021-02-11.1	&	11 04	&	-19 03	&	1.098	&	0.139	&	34.7	&	OWL MAR 				&	$R$ \\
2021-02-11.2	&	11 04	&	-19 02	&	1.098	&	0.139	&	34.6	&	LCO CTIO A 				&	$R$ \\
2021-02-11.3	&	11 04	&	-19 01	&	1.098	&	0.139	&	34.5	&	LCO CTIO B 				&	$R$ \\
2021-02-11.3	&	11 04	&	-19 01	&	1.098	&	0.139	&	34.5	&	Skynet Prompt5 			&	$R$ \\
2021-02-11.4	&	11 03	&	-19 00	&	1.098	&	0.139	&	34.3	&	Winer Observatory RBT 	&	$R$ \\
2021-02-11.5	&	11 03	&	-18 59	&	1.098	&	0.138	&	34.2	&	LCO SSO A 				&	$R$ \\
2021-02-11.5	&	11 03	&	-18 59	&	1.098	&	0.138	&	34.2	&	OWL USA 				&	$R$ \\
2021-02-12.0	&	11 01	&	-18 54	&	1.099	&	0.137	&	33.7	&	Adiyaman Observatory 	&	$R$ \\
2021-02-12.2	&	11 01	&	-18 52	&	1.099	&	0.137	&	33.4	&	OWL MAR 				&	$R$ \\
2021-02-12.3	&	11 00	&	-18 51	&	1.099	&	0.137	&	33.3	&	Winer Observatory RBT 	&	$R$ \\
2021-02-12.4	&	11 00	&	-18 50	&	1.099	&	0.136	&	33.2	&	OWL USA 				&	$R$ \\
2021-02-12.7	&	10 59	&	-18 46	&	1.099	&	0.136	&	32.9	&	Skynet Prompt MO1 		&	$R$ \\
2021-02-13.2	&	10 57	&	-18 40	&	1.099	&	0.135	&	32.3	&	Skynet Prompt5 			&	$R$ \\
2021-02-14.0	&	10 54	&	-18 29	&	1.099	&	0.133	&	31.4	&	Adiyaman Observatory 	&	$R$ \\
2021-02-15.1	&	10 50	&	-18 13	&	1.099	&	0.131	&	30.1	&	CAHA 2.2 m 				&	$R$ \\
2021-02-16.1	&	10 47	&	-17 55	&	1.099	&	0.129	&	29.0	&	OWL MAR					&	$R$ \\
2021-02-16.4	&	10 45	&	-17 50	&	1.099	&	0.129	&	28.6	&	Skynet Prompt5 			&	$R$ \\
2021-02-16.6	&	10 45	&	-17 46	&	1.099	&	0.128	&	28.4	&	Skynet Prompt MO1 		&	$R$ \\
2021-02-16.9	&	10 43	&	-17 40	&	1.099	&	0.128	&	28.1	&	OWL MAR 				&	$R$ \\
2021-02-17.4	&	10 41	&	-17 30	&	1.099	&	0.127	&	27.5	&	Winer Observatory RBT 	&	$R$ \\
2021-02-17.7	&	10 40	&	-17 24	&	1.099	&	0.127	&	27.2	&	Skynet Prompt MO1 		&	$R$ \\
2021-02-18.4	&	10 37	&	-17 09	&	1.099	&	0.125	&	26.5	&	LOAO 					&	$R$ \\
2021-02-18.4	&	10 37	&	-17 09	&	1.099	&	0.125	&	26.5	&	Winer Observatory RBT 	&	$R$ \\
2021-02-18.7	&	10 36	&	-17 02	&	1.099	&	0.125	&	26.1	&	SOAO 					&	$R$ \\
2021-02-18.8	&	10 36	&	-16 60	&	1.099	&	0.125	&	26.0	&	Skynet Prompt MO1 		&	$R$ \\
2021-02-19.0	&	10 35	&	-16 55	&	1.099	&	0.125	&	25.8	&	Suhora Observatory 		&	$R$ \\
2021-02-19.3	&	10 34	&	-16 48	&	1.099	&	0.124	&	25.5	&	LOAO 					&	$R$ \\
2021-02-19.3	&	10 34	&	-16 48	&	1.099	&	0.124	&	25.5	&	Winer Observatory RBT 	&	$R$ \\
2021-02-19.6	&	10 32	&	-16 40	&	1.099	&	0.124	&	25.2	&	BOAO 					&	$R$ \\
2021-02-19.6	&	10 32	&	-16 40	&	1.099	&	0.124	&	25.2	&	OWL KOR 				&	$R$ \\
2021-02-19.6	&	10 32	&	-16 40	&	1.099	&	0.124	&	25.2	&	SOAO 					&	$R$ \\
2021-02-19.7	&	10 32	&	-16 38	&	1.099	&	0.123	&	25.2	&	Skynet Prompt MO1 		&	$R$ \\
2021-02-20.3	&	10 29	&	-16 23	&	1.099	&	0.123	&	24.6	&	LOAO 					&	$R$ \\
2021-02-20.5	&	10 28	&	-16 17	&	1.099	&	0.122	&	24.4	&	Skynet Prompt MO1 		&	$R$ \\
2021-02-20.6	&	10 28	&	-16 15	&	1.099	&	0.122	&	24.3	&	BOAO 					&	$R$ \\
2021-02-21.0	&	10 26	&	-16 04	&	1.099	&	0.122	&	24.0	&	Suhora Observatory 		&	$R$ \\
2021-02-21.4	&	10 25	&	-15 52	&	1.099	&	0.121	&	23.6	&	Winer Observatory RBT 	&	$R$ \\
2021-02-21.6	&	10 24	&	-15 47	&	1.099	&	0.121	&	23.5	&	BOAO 					&	$R$ \\
2021-02-21.8	&	10 23	&	-15 41	&	1.099	&	0.121	&	23.3	&	OWL ISR 				&	$R$ \\
2021-02-22.5	&	10 20	&	-15 20	&	1.099	&	0.120	&	22.8	&	Skynet Prompt MO1 		&	$R$ \\
2021-02-22.7	&	10 19	&	-15 14	&	1.098	&	0.119	&	22.6	&	BOAO 					&	$R$ \\
2021-02-22.7	&	10 19	&	-15 14	&	1.098	&	0.119	&	22.6	&	Skynet Prompt MO1 		&	$R$ \\
2021-02-23.1	&	10 17	&	-15 01	&	1.098	&	0.119	&	22.4	&	OWL USA 				&	$R$ \\
2021-02-24.2	&	10 12	&	-14 25	&	1.098	&	0.118	&	21.8	&	OWL USA 				&	$R$ \\
2021-02-25.0	&	10 08	&	-13 58	&	1.098	&	0.117	&	21.4	&	Suhora Observatory 		&	$R$ \\
2021-02-26.2	&	10 03	&	-13 14	&	1.097	&	0.116	&	21.1	&	OWL USA 				&	$R$ \\
2021-02-27.2	&	09 58	&	-12 36	&	1.097	&	0.115	&	21.1	&	OWL USA 				&	$R$ \\
2021-02-28.0	&	09 54	&	-12 04	&	1.096	&	0.115	&	21.2	&	CAHA 2.2 m 				&	$R$ \\
2021-02-28.0	&	09 54	&	-12 04	&	1.096	&	0.115	&	21.2	&	Skynet Prompt5 			&	$R$ \\
2021-03-01.2	&	09 49	&	-11 14	&	1.096	&	0.114	&	21.5	&	OWL USA 				&	$R$ \\
2021-03-01.9	&	09 45	&	-10 45	&	1.095	&	0.114	&	21.9	&	TUG 					&	$R$ \\
2021-03-03.0	&	09 40	&	-09 57	&	1.095	&	0.113	&	22.6	&	OWL MAR 				&	$R$ \\
2021-03-03.0	&	09 40	&	-09 57	&	1.095	&	0.113	&	22.6	&	Suhora Observatory 		&	$R$ \\
2021-03-03.0	&	09 40	&	-09 57	&	1.095	&	0.113	&	22.6	&	KMTNet CTIO 			&	$VR$ \\
2021-03-03.0	&	09 40	&	-09 57	&	1.095	&	0.113	&	22.6	&	Skynet Prompt5 			&	$R$ \\
2021-03-03.2	&	09 39	&	-09 48	&	1.095	&	0.113	&	22.7	&	OWL USA 				&	$R$ \\
2021-03-03.2	&	09 39	&	-09 48	&	1.095	&	0.113	&	22.7	&	Winer Observatory RBT 	&	$R$ \\
2021-03-03.4	&	09 39	&	-09 39	&	1.095	&	0.113	&	22.9	&	KMTNet SSO 				&	$VR$ \\
2021-03-05.1	&	09 31	&	-08 21	&	1.093	&	0.113	&	24.4	&	LCO CTIO A 				&	$R$ \\
2021-03-05.2	&	09 30	&	-08 16	&	1.093	&	0.113	&	24.5	&	LCO McDonald A 			&	$R$ \\
2021-03-05.4	&	09 30	&	-08 07	&	1.093	&	0.113	&	24.7	&	LCO SSO B 				&	$R$ \\
2021-03-05.6	&	09 29	&	-07 58	&	1.093	&	0.113	&	24.9	&	LCO SSO A 				&	$R$ \\
2021-03-05.8	&	09 28	&	-07 48	&	1.093	&	0.113	&	25.1	&	LCO SAAO A 				&	$R$ \\
2021-03-05.8	&	09 28	&	-07 48	&	1.093	&	0.113	&	25.1	&	LCO SAAO C 				&	$R$ \\
2021-03-06.4	&	09 25	&	-07 20	&	1.092	&	0.113	&	25.8	&	LCO SSO B 				&	$R$ \\
2021-03-06.5	&	09 25	&	-07 15	&	1.092	&	0.113	&	25.9	&	LCO SSO A 				&	$R$ \\
2021-03-06.8	&	09 23	&	-07 01	&	1.092	&	0.113	&	26.3	&	LCO SAAO C 				&	$R$ \\
2021-03-06.8	&	09 23	&	-07 01	&	1.092	&	0.113	&	26.3	&	OWL ISR 				&	$R$ \\
2021-03-06.9	&	09 23	&	-06 56	&	1.092	&	0.113	&	26.4	&	LCO SAAO A 				&	$R$ \\
2021-03-06.9	&	09 23	&	-06 56	&	1.092	&	0.113	&	26.4	&	OAUJ CDK500 			&	$R$ \\
2021-03-07.0	&	09 22	&	-06 51	&	1.092	&	0.113	&	26.5	&	LCO CTIO B				&	$R$ \\
2021-03-07.1	&	09 22	&	-06 47	&	1.092	&	0.113	&	26.6	&	OWL USA 				&	$R$ \\
2021-03-07.2	&	09 22	&	-06 42	&	1.092	&	0.113	&	26.8	&	LCO McDonald A 			&	$R$ \\
2021-03-07.3	&	09 21	&	-06 37	&	1.092	&	0.113	&	26.9	&	LCO McDonald B 			&	$R$ \\
2021-03-07.4	&	09 21	&	-06 32	&	1.092	&	0.113	&	27.0	&	KMTNet SSO 				&	$VR$ \\
2021-03-07.5	&	09 20	&	-06 27	&	1.091	&	0.113	&	27.1	&	LCO SSO A 				&	$R$ \\
2021-03-07.6	&	09 20	&	-06 23	&	1.091	&	0.113	&	27.3	&	LCO SSO B 				&	$R$ \\
2021-03-07.8	&	09 19	&	-06 13	&	1.091	&	0.113	&	27.5	&	KMTNet SAAO 			&	$VR$ \\
2021-03-07.8	&	09 19	&	-06 13	&	1.091	&	0.113	&	27.5	&	LCO SAAO A 				&	$R$ \\
2021-03-07.9	&	09 19	&	-06 08	&	1.091	&	0.113	&	27.6	&	LCO SAAO C 				&	$R$ \\
2021-03-08.0	&	09 18	&	-06 03	&	1.091	&	0.113	&	27.8	&	LCO CTIO B 				&	$R$ \\
2021-03-08.1	&	09 18	&	-05 58	&	1.091	&	0.113	&	27.9	&	LCO McDonald A 			&	$R$ \\
2021-03-08.2	&	09 17	&	-05 54	&	1.091	&	0.113	&	28.0	&	LCO McDonald B 			&	$R$ \\
2021-03-08.8	&	09 15	&	-05 25	&	1.090	&	0.113	&	28.8	&	OWL ISR 				&	$R$ \\
2021-03-08.9	&	09 14	&	-05 20	&	1.090	&	0.113	&	29.0	&	LCO SAAO C				&	$R$ \\
2021-03-09.0	&	09 14	&	-05 15	&	1.090	&	0.113	&	29.1	&	LCO CTIO B				&	$R$ \\
2021-03-09.0	&	09 14	&	-05 15	&	1.090	&	0.113	&	29.1	&	LCO SAAO B				&	$R$ \\
2021-03-09.5	&	09 12	&	-04 51	&	1.090	&	0.113	&	29.8	&	LCO SSO A 				&	$R$ \\
2021-03-09.6	&	09 12	&	-04 46	&	1.090	&	0.113	&	29.9	&	LCO SSO B 				&	$R$ \\
2021-03-09.6	&	09 12	&	-04 46	&	1.090	&	0.113	&	29.9	&	SOAO 					&	$R$ \\
2021-03-09.8	&	09 11	&	-04 36	&	1.089	&	0.113	&	30.2	&	KMTNet SAAO 			&	$VR$ \\
2021-03-09.8	&	09 11	&	-04 36	&	1.089	&	0.113	&	30.2	&	LCO SAAO B 				&	$R$ \\
2021-03-10.0	&	09 10	&	-04 26	&	1.089	&	0.113	&	30.5	&	Suhora Observatory 		&	$R$ \\
2021-03-10.2	&	09 09	&	-04 16	&	1.089	&	0.114	&	30.8	&	LCO CTIO A 				&	$R$ \\
2021-03-10.2	&	09 09	&	-04 16	&	1.089	&	0.114	&	30.8	&	Skynet DSO-14 			&	$R$ \\
2021-03-10.3	&	09 09	&	-04 12	&	1.089	&	0.114	&	30.9	&	OWL USA 				&	$R$ \\
2021-03-10.3	&	09 09	&	-04 12	&	1.089	&	0.114	&	30.9	&	Skynet RRRT 			&	$R$ \\
2021-03-10.3	&	09 09	&	-04 12	&	1.089	&	0.114	&	30.9	&	Winer Observatory RBT 	&	$R$ \\
2021-03-10.5	&	09 08	&	-04 02	&	1.089	&	0.114	&	31.2	&	BOAO 					&	$R$ \\
2021-03-10.5	&	09 08	&	-04 02	&	1.089	&	0.114	&	31.2	&	DOAO 					&	$R$ \\
2021-03-10.6	&	09 08	&	-03 57	&	1.089	&	0.114	&	31.3	&	SOAO 					&	$R$ \\
2021-03-10.6	&	09 08	&	-03 57	&	1.089	&	0.114	&	31.3	&	Kawabe Observatory 		&	$r'$ \\
2021-03-10.9	&	09 06	&	-03 42	&	1.088	&	0.114	&	31.8	&	CAHA 1.23 m 			&	$R$ \\
2021-03-10.9	&	09 06	&	-03 42	&	1.088	&	0.114	&	31.8	&	OAUJ CDK500 			&	$R$ \\
2021-03-11.0	&	09 06	&	-03 37	&	1.088	&	0.114	&	31.9	&	Suhora Observatory 		&	$R$ \\
2021-03-11.1	&	09 06	&	-03 33	&	1.088	&	0.114	&	32.1	&	LCO CTIO B 				&	$R$ \\
2021-03-11.2	&	09 05	&	-03 28	&	1.088	&	0.114	&	32.2	&	Skynet RRRT 			&	$R$ \\
2021-03-11.3	&	09 05	&	-03 23	&	1.088	&	0.114	&	32.4	&	OWL USA 				&	$R$ \\
2021-03-11.3	&	09 05	&	-03 23	&	1.088	&	0.114	&	32.4	&	Winer Observatory RBT 	&	$R$ \\
2021-03-11.8	&	09 03	&	-02 59	&	1.087	&	0.114	&	33.1	&	LCO SAAO A 				&	$R$ \\
2021-03-11.9	&	09 02	&	-02 54	&	1.087	&	0.114	&	33.2	&	LCO SAAO B				&	$R$ \\
2021-03-11.9	&	09 02	&	-02 54	&	1.087	&	0.114	&	33.2	&	LCO SAAO C 				&	$R$ \\
2021-03-12.0	&	09 02	&	-02 49	&	1.087	&	0.114	&	33.4	&	LCO CTIO B 				&	$R$ \\
2021-03-12.2	&	09 01	&	-02 39	&	1.087	&	0.115	&	33.7	&	LCO CTIO A 				&	$R$ \\
2021-03-12.3	&	09 01	&	-02 34	&	1.087	&	0.115	&	33.8	&	Winer Observatory RBT 	&	$R$ \\
2021-03-12.6	&	08 60	&	-02 20	&	1.086	&	0.115	&	34.3	&	LCO SSO A 				&	$R$ \\
2021-03-12.7	&	08 60	&	-02 15	&	1.086	&	0.115	&	34.4	&	LCO SAAO B 				&	$R$ \\
2021-03-12.9	&	08 59	&	-02 05	&	1.086	&	0.115	&	34.7	&	LCO SAAO A 				&	$R$ \\
2021-03-12.9	&	08 59	&	-02 05	&	1.086	&	0.115	&	34.7	&	OAUJ CDK500 			&	$R$ \\
2021-03-12.9	&	08 59	&	-02 05	&	1.086	&	0.115	&	34.7	&	Suhora Observatory 		&	$R$ \\
2021-03-13.2	&	08 58	&	-01 51	&	1.086	&	0.115	&	35.2	&	LCO CTIO B 				&	$R$ \\
2021-03-13.6	&	08 56	&	-01 36	&	1.086	&	0.115	&	35.6	&	DOAO 					&	$R$ \\
2021-03-13.8	&	08 56	&	-01 22	&	1.085	&	0.116	&	36.1	&	TUG 					&	$R$ \\
2021-03-13.9	&	08 55	&	-01 17	&	1.085	&	0.116	&	36.2	&	LCO SAAO A 				&	$R$ \\
2021-03-14.0	&	08 55	&	-01 12	&	1.085	&	0.116	&	36.4	&	Skynet Prompt5 			&	$R$ \\
2021-03-14.0	&	08 55	&	-01 12	&	1.085	&	0.116	&	36.4	&	Skynet RRRT 			&	$R$ \\
2021-03-14.1	&	08 54	&	-01 07	&	1.085	&	0.116	&	36.5	&	LCO CTIO B 				&	$R$ \\
2021-03-14.3	&	08 54	&	+00 58	&	1.085	&	0.116	&	36.8	&	LCO McDonald A 			&	$R$ \\
2021-03-14.5	&	08 53	&	+00 48	&	1.084	&	0.116	&	37.1	&	BOAO 					&	$R$ \\
2021-03-15.0	&	08 51	&	+00 25	&	1.084	&	0.116	&	37.9	&	OAUJ CDK500 			&	$R$ \\
2021-03-15.0	&	08 51	&	+00 25	&	1.084	&	0.116	&	37.9	&	Skynet Prompt5 			&	$R$ \\
2021-03-15.1	&	08 51	&	+00 20	&	1.084	&	0.117	&	38.0	&	Skynet Prompt6 			&	$R$ \\
2021-03-15.1	&	08 51	&	+00 20	&	1.084	&	0.117	&	38.0	&	Skynet RRRT 			&	$R$ \\
2021-03-15.2	&	08 51	&	+00 15	&	1.084	&	0.117	&	38.2	&	LOAO 					&	$R$ \\
2021-03-15.9	&	08 48	&	+00 18	&	1.083	&	0.117	&	39.3	&	OAUJ CDK500 			&	$R$ \\
2021-03-16.0	&	08 48	&	+00 22	&	1.083	&	0.117	&	39.4	&	CAHA 2.2 m 				&	$R$ \\
2021-03-16.2	&	08 47	&	+00 32	&	1.082	&	0.118	&	39.7	&	LOAO 					&	$R$ \\
2021-03-16.6	&	08 46	&	+00 50	&	1.082	&	0.118	&	40.3	&	BOAO 					&	$R$ \\
2021-03-16.6	&	08 46	&	+00 50	&	1.082	&	0.118	&	40.3	&	DOAO 					&	$R$ \\
2021-03-16.9	&	08 45	&	+01 04	&	1.081	&	0.118	&	40.8	&	Adiyaman Observatory 	&	$R$ \\
2021-03-17.0	&	08 45	&	+01 09	&	1.081	&	0.118	&	40.9	&	CAHA 2.2 m 				&	$R$ \\
2021-03-17.0	&	08 45	&	+01 09	&	1.081	&	0.118	&	40.9	&	Skynet Prompt5 			&	$R$ \\
2021-03-17.1	&	08 45	&	+01 13	&	1.081	&	0.118	&	41.1	&	Skynet Prompt6 			&	$R$ \\
2021-03-17.6	&	08 43	&	+01 32	&	1.081	&	0.119	&	41.7	&	DOAO 					&	$R$ \\
2021-03-18.0	&	08 42	&	+01 55	&	1.080	&	0.119	&	42.4	&	CAHA 2.2 m 				&	$R$ \\
2021-03-18.2	&	08 41	&	+02 04	&	1.080	&	0.119	&	42.7	&	OWL USA 				&	$R$ \\
2021-03-18.4	&	08 41	&	+02 13	&	1.079	&	0.120	&	43.0	&	KMTNet SSO 				&	$VR$ \\
2021-03-18.8	&	08 40	&	+02 31	&	1.079	&	0.120	&	43.6	&	KMTNet SAAO 			&	$VR$ \\
2021-03-18.8	&	08 40	&	+02 31	&	1.079	&	0.120	&	43.6	&	OWL ISR 				&	$R$ \\
2021-03-19.2	&	08 38	&	+02 48	&	1.078	&	0.120	&	44.2	&	OWL USA 				&	$R$ \\
2021-03-19.9	&	08 37	&	+03 19	&	1.077	&	0.121	&	45.2	&	OWL ISR 				&	$R$ \\
2021-03-20.2	&	08 36	&	+03 32	&	1.077	&	0.122	&	45.7	&	OWL USA					&	$R$ \\
2021-03-20.2	&	08 36	&	+03 32	&	1.077	&	0.122	&	45.7	&	Skynet RRRT 			&	$R$ \\
2021-03-21.2	&	08 33	&	+04 15	&	1.076	&	0.123	&	47.1	&	OWL USA 				&	$R$ \\
2021-03-22.3	&	08 30	&	+05 01	&	1.074	&	0.124	&	48.7	&	OWL USA 				&	$R$ \\
2021-03-22.5	&	08 30	&	+05 10	&	1.074	&	0.124	&	49.0	&	DOAO 					&	$R$ \\
2021-03-22.8	&	08 29	&	+05 22	&	1.073	&	0.125	&	49.4	&	KMTNet SAAO				&	$VR$ \\
2021-03-24.8	&	08 25	&	+06 42	&	1.070	&	0.127	&	52.2	&	KMTNet SAAO				&	$VR$ \\
2021-03-30.0	&	08 16	&	+09 53	&	1.061	&	0.134	&	59.0	&	KMTNet CTIO				&	$VR$ \\
2021-03-30.7	&	08 15	&	+10 16	&	1.060	&	0.135	&	59.9	&	KMTNet SAAO				&	$VR$ \\
2021-04-01.8	&	08 12	&	+11 25	&	1.055	&	0.138	&	62.5	&	KMTNet SAAO				&	$VR$ \\
2021-04-03.1	&	08 11	&	+12 06	&	1.053	&	0.139	&	64.1	&	KMTNet CTIO				&	$VR$ \\
2021-04-04.9	&	08 09	&	+13 00	&	1.049	&	0.142	&	66.2	&	CAHA 1.23 m				&	$R$ \\
2021-04-05.4	&	08 08	&	+13 15	&	1.048	&	0.142	&	66.7	&	KMTNet SSO 				&	$VR$ \\
2021-04-05.9	&	08 08	&	+13 30	&	1.047	&	0.143	&	67.3	&	CAHA 1.23 m				&	$R$ \\
2021-04-08.4	&	08 06	&	+14 39	&	1.042	&	0.146	&	70.1	&	KMTNet SSO 				&	$VR$ \\
2021-04-08.8	&	08 06	&	+14 50	&	1.041	&	0.147	&	70.6	&	KMTNet SAAO				&	$VR$ \\
2021-04-14.8	&	08 02	&	+17 22	&	1.026	&	0.154	&	77.1	&	KMTNet SAAO				&	$VR$ \\
2021-04-16.8	&	08 02	&	+18 09	&	1.021	&	0.156	&	79.2	&	KMTNet SAAO				&	$VR$ \\
2021-04-18.4	&	08 01	&	+18 45	&	1.017	&	0.158	&	80.8	&	KMTNet SSO 				&	$VR$ \\
2021-04-18.8	&	08 01	&	+18 54	&	1.016	&	0.158	&	81.2	&	KMTNet SAAO				&	$VR$ \\
2021-04-20.4	&	08 01	&	+19 29	&	1.012	&	0.160	&	82.9	&	KMTNet SSO 				&	$VR$ \\
\multicolumn{7}{c}{{\bf –Sparse Photometry-}} \\
\multicolumn{3}{l}{2020-11-12 2021-04-09}  	      		  & 	  & 	  & 	 & ATLAS & $c$ \\
\multicolumn{3}{l}{2020-12-01 2021-04-03} 	      		  & 	  & 	  & 	 & ATLAS & $o$ \\
\multicolumn{7}{c}{{\bf –Space-based Photometry-}} \\
\multicolumn{3}{l}{2021-02-19 2021-03-07} 	      		  & 	  & 	  & 	 & TESS & wide $I$ \\

\end{longtable}

\tablefoot{R. A. : Right ascension, Dec. : Declination, $\Delta$ : geocentric distance, $r_{h}$ : heliocentric distance, $\alpha$ : phase angle.}

\newpage

\begin{longtable}{lllllll}
\caption{List of the historical light curves.}
\label{table:A2}\\
\hline\hline
            Date UT   & R. A. & Dec.  & $r_{h}$    &  $\Delta$ & $\alpha$    & References \\
                      & (h m) & ($^\circ$ ')  & (AU)  & (AU)  & ($^\circ$)  &       \\
\hline
\endfirsthead
\caption{continued.}\\
\hline\hline
            Date UT   & R. A. & Dec.  & $r_{h}$    &  $\Delta$ & $\alpha$    & References \\
                      & (h m) & ($^\circ$ ')  & (AU)  & (AU)  & ($^\circ$)  &       \\
\hline
\endhead
\hline
\endfoot
2012-12-23.3	&	10 42	&	-27 22	&	1.000	&	0.102	&	77.6	&	\cite{Pravec_et_al_2014} \\
2012-12-25.3	&	10 33	&	-27 23	&	1.006	&	0.101	&	74.3	&	\cite{Pravec_et_al_2014} \\
2012-12-26.3	&	10 28	&	-27 23	&	1.009	&	0.101	&	72.7	&	\cite{Pravec_et_al_2014} \\
2012-12-27.2	&	10 23	&	-27 21	&	1.011	&	0.101	&	71.2	&	\cite{Pravec_et_al_2014} \\
2012-12-28.2	&	10 18	&	-27 19	&	1.014	&	0.100	&	69.5	&	\cite{Pravec_et_al_2014} \\
2012-12-29.2	&	10 13	&	-27 14	&	1.017	&	0.100	&	67.8	&	\cite{Pravec_et_al_2014} \\
2012-12-30.2	&	10 08	&	-27 10	&	1.019	&	0.099	&	66.2	&	\cite{Pravec_et_al_2014} \\
2012-12-31.2	&	10 03	&	-27 03	&	1.022	&	0.099	&	64.5	&	\cite{Pravec_et_al_2014} \\
2013-01-03.2	&	09 47	&	-26 34	&	1.029	&	0.098	&	59.5	&	\cite{Pravec_et_al_2014} \\
2013-01-04.2	&	09 41	&	-26 20	&	1.032	&	0.097	&	57.9	&	\cite{Pravec_et_al_2014} \\
2013-01-05.2	&	09 36	&	-26 06	&	1.034	&	0.097	&	56.2	&	\cite{Pravec_et_al_2014} \\
2013-01-06.1	&	09 30	&	-25 51	&	1.036	&	0.097	&	54.7	&	\cite{Pravec_et_al_2014} \\
2013-01-06.2	&	09 30	&	-25 49	&	1.036	&	0.097	&	54.6	&	\cite{Pravec_et_al_2014} \\
2013-01-07.2	&	09 24	&	-25 30	&	1.039	&	0.097	&	52.9	&	\cite{Pravec_et_al_2014} \\
2013-01-07.6	&	09 22	&	-25 22	&	1.040	&	0.097	&	52.3	&	\cite{Pravec_et_al_2014} \\
2013-01-08.1	&	09 19	&	-25 11	&	1.041	&	0.097	&	51.5	&	\cite{Pravec_et_al_2014} \\
2013-01-08.1	&	09 19	&	-25 11	&	1.041	&	0.097	&	51.5	&	\cite{Pravec_et_al_2014} \\
2013-01-08.2	&	09 18	&	-25 10	&	1.041	&	0.097	&	51.3	&	\cite{Pravec_et_al_2014} \\
2013-01-09.1	&	09 13	&	-24 49	&	1.043	&	0.097	&	49.9	&	\cite{Pravec_et_al_2014} \\
2013-01-09.1	&	09 13	&	-24 49	&	1.043	&	0.097	&	49.9	&	\cite{Pravec_et_al_2014} \\
2013-01-09.2	&	09 12	&	-24 46	&	1.043	&	0.097	&	49.7	&	\cite{Pravec_et_al_2014} \\
2013-01-09.3	&	09 12	&	-24 44	&	1.043	&	0.097	&	49.5	&	\cite{Pravec_et_al_2014} \\
2013-01-10.1	&	09 07	&	-24 24	&	1.045	&	0.097	&	48.3	&	\cite{Pravec_et_al_2014} \\
2013-01-10.2	&	09 06	&	-24 22	&	1.045	&	0.097	&	48.1	&	\cite{Pravec_et_al_2014} \\
2013-01-10.3	&	09 06	&	-24 19	&	1.046	&	0.097	&	48.0	&	\cite{Pravec_et_al_2014} \\
2013-01-11.1	&	09 01	&	-23 57	&	1.047	&	0.097	&	46.7	&	\cite{Pravec_et_al_2014} \\
2013-01-11.3	&	09 00	&	-23 52	&	1.048	&	0.097	&	46.4	&	\cite{Pravec_et_al_2014} \\
2013-01-11.5	&	08 59	&	-23 46	&	1.048	&	0.097	&	46.1	&	\cite{Pravec_et_al_2014} \\
2013-01-12.1	&	08 55	&	-23 28	&	1.049	&	0.097	&	45.2	&	\cite{Pravec_et_al_2014} \\
2013-01-12.2	&	08 55	&	-23 25	&	1.050	&	0.097	&	45.1	&	\cite{Pravec_et_al_2014} \\
2013-01-12.2	&	08 55	&	-23 25	&	1.050	&	0.097	&	45.1	&	\cite{Pravec_et_al_2014} \\
2013-01-13.1	&	08 49	&	-22 57	&	1.051	&	0.097	&	43.7	&	\cite{Pravec_et_al_2014} \\
2013-01-13.2	&	08 49	&	-22 54	&	1.052	&	0.097	&	43.6	&	\cite{Pravec_et_al_2014} \\
2013-01-14.1	&	08 44	&	-22 24	&	1.053	&	0.097	&	42.3	&	\cite{Pravec_et_al_2014} \\
2013-01-14.1	&	08 44	&	-22 24	&	1.053	&	0.097	&	42.3	&	\cite{Pravec_et_al_2014} \\
2013-01-14.2	&	08 43	&	-22 20	&	1.054	&	0.097	&	42.2	&	\cite{Pravec_et_al_2014} \\
2013-01-15.1	&	08 38	&	-21 49	&	1.055	&	0.098	&	40.9	&	\cite{Pravec_et_al_2014} \\
2013-01-15.2	&	08 37	&	-21 45	&	1.056	&	0.098	&	40.8	&	\cite{Pravec_et_al_2014} \\
2013-01-15.5	&	08 35	&	-21 34	&	1.056	&	0.098	&	40.4	&	\cite{Pravec_et_al_2014} \\
2013-01-16.1	&	08 32	&	-21 12	&	1.057	&	0.098	&	39.6	&	\cite{Pravec_et_al_2014} \\
2013-01-16.1	&	08 32	&	-21 12	&	1.057	&	0.098	&	39.6	&	\cite{Pravec_et_al_2014} \\
2013-01-16.3	&	08 31	&	-21 04	&	1.058	&	0.098	&	39.4	&	\cite{Pravec_et_al_2014} \\
2013-01-16.3	&	08 31	&	-21 04	&	1.058	&	0.098	&	39.4	&	\cite{Pravec_et_al_2014} \\
2013-01-19.1	&	08 15	&	-19 11	&	1.063	&	0.100	&	36.2	&	\cite{Pravec_et_al_2014} \\
2013-01-19.2	&	08 15	&	-19 07	&	1.063	&	0.100	&	36.1	&	\cite{Pravec_et_al_2014} \\
2013-01-20.1	&	08 10	&	-18 28	&	1.065	&	0.101	&	35.2	&	\cite{Pravec_et_al_2014} \\
2013-01-20.2	&	08 10	&	-18 24	&	1.065	&	0.101	&	35.1	&	\cite{Pravec_et_al_2014} \\
2013-01-22.2	&	07 59	&	-16 54	&	1.068	&	0.103	&	33.5	&	\cite{Pravec_et_al_2014} \\
2013-01-22.3	&	07 59	&	-16 50	&	1.068	&	0.103	&	33.5	&	\cite{Pravec_et_al_2014} \\
2013-01-23.1	&	07 55	&	-16 13	&	1.070	&	0.104	&	33.0	&	\cite{Pravec_et_al_2014} \\
2013-01-23.2	&	07 55	&	-16 08	&	1.070	&	0.104	&	32.9	&	\cite{Pravec_et_al_2014} \\
2013-01-24.1	&	07 50	&	-15 26	&	1.071	&	0.105	&	32.4	&	\cite{Pravec_et_al_2014} \\
2013-01-24.2	&	07 50	&	-15 22	&	1.071	&	0.105	&	32.4	&	\cite{Pravec_et_al_2014} \\
2013-01-24.3	&	07 49	&	-15 17	&	1.072	&	0.105	&	32.4	&	\cite{Pravec_et_al_2014} \\
2013-02-04.1	&	07 10	&	-06 46	&	1.086	&	0.123	&	33.8	&	\cite{Pravec_et_al_2014} \\
2013-02-05.1	&	07 07	&	-06 01	&	1.087	&	0.125	&	34.4	&	\cite{Pravec_et_al_2014} \\
2013-02-05.3	&	07 07	&	-05 53	&	1.087	&	0.126	&	34.5	&	\cite{Pravec_et_al_2014} \\
2013-02-06.1	&	07 05	&	-05 17	&	1.088	&	0.128	&	35.0	&	\cite{Pravec_et_al_2014} \\
2013-02-06.2	&	07 04	&	-05 13	&	1.088	&	0.128	&	35.1	&	\cite{Pravec_et_al_2014} \\
2013-02-06.9	&	07 03	&	-04 43	&	1.089	&	0.129	&	35.5	&	\cite{Pravec_et_al_2014} \\
2013-02-07.1	&	07 02	&	-04 34	&	1.089	&	0.130	&	35.7	&	\cite{Pravec_et_al_2014} \\
2013-02-07.2	&	07 02	&	-04 29	&	1.089	&	0.130	&	35.7	&	\cite{Pravec_et_al_2014} \\
2013-02-08.1	&	07 00	&	-03 51	&	1.090	&	0.132	&	36.4	&	\cite{Pravec_et_al_2014} \\
2013-02-08.2	&	07 00	&	-03 47	&	1.090	&	0.132	&	36.4	&	\cite{Pravec_et_al_2014} \\
2013-02-12.1	&	06 53	&	-01 09	&	1.093	&	0.142	&	39.3	&	\cite{Pravec_et_al_2014} \\
2013-02-13.1	&	06 52	&	-00 31	&	1.094	&	0.145	&	40.1	&	\cite{Pravec_et_al_2014} \\
2013-02-13.2	&	06 52	&	-00 27	&	1.094	&	0.145	&	40.2	&	\cite{Pravec_et_al_2014} \\
2013-02-14.2	&	06 50	&	+00 10	&	1.094	&	0.148	&	40.9	&	\cite{Pravec_et_al_2014} \\
2013-02-14.2	&	06 50	&	+00 10	&	1.094	&	0.148	&	40.9	&	\cite{Pravec_et_al_2014} \\
2013-02-15.1	&	06 49	&	+00 43	&	1.095	&	0.150	&	41.6	&	\cite{Pravec_et_al_2014} \\
2013-02-15.1	&	06 49	&	+00 43	&	1.095	&	0.150	&	41.6	&	\cite{Pravec_et_al_2014} \\
2013-02-16.0	&	06 48	&	+01 15	&	1.095	&	0.152	&	42.3	&	\cite{Pravec_et_al_2014} \\
2013-02-16.2	&	06 48	&	+01 22	&	1.096	&	0.153	&	42.5	&	\cite{Pravec_et_al_2014} \\
2013-02-16.2	&	06 48	&	+01 22	&	1.096	&	0.153	&	42.5	&	\cite{Pravec_et_al_2014} \\
2013-02-17.0	&	06 48	&	+01 49	&	1.096	&	0.155	&	43.0	&	\cite{Pravec_et_al_2014} \\
2013-02-17.2	&	06 47	&	+01 56	&	1.096	&	0.156	&	43.2	&	\cite{Pravec_et_al_2014} \\
2013-02-17.2	&	06 47	&	+01 56	&	1.096	&	0.156	&	43.2	&	\cite{Pravec_et_al_2014} \\
2013-02-18.0	&	06 47	&	+02 23	&	1.096	&	0.158	&	43.8	&	\cite{Pravec_et_al_2014} \\
2013-02-18.0	&	06 47	&	+02 23	&	1.096	&	0.158	&	43.8	&	\cite{Pravec_et_al_2014} \\
2013-02-18.2	&	06 47	&	+02 30	&	1.096	&	0.159	&	44.0	&	\cite{Pravec_et_al_2014} \\
2013-02-19.9	&	06 46	&	+03 25	&	1.097	&	0.163	&	45.2	&	\cite{Pravec_et_al_2014} \\
2013-03-09.1	&	06 49	&	+10 38	&	1.096	&	0.215	&	56.3	&	\cite{Pravec_et_al_2014} \\
2013-03-10.2	&	06 50	&	+11 00	&	1.095	&	0.218	&	56.9	&	\cite{Pravec_et_al_2014} \\
2013-03-11.0	&	06 51	&	+11 15	&	1.095	&	0.221	&	57.3	&	\cite{Pravec_et_al_2014} \\
2013-03-12.0	&	06 52	&	+11 33	&	1.094	&	0.224	&	57.9	&	\cite{Pravec_et_al_2014} \\
2013-03-13.0	&	06 53	&	+11 51	&	1.094	&	0.227	&	58.4	&	\cite{Pravec_et_al_2014} \\
2013-04-09.0	&	07 33	&	+17 20	&	1.060	&	0.297	&	70.7	&	\cite{Pravec_et_al_2014} \\
2013-04-12.0	&	07 38	&	+17 43	&	1.054	&	0.303	&	71.9	&	\cite{Pravec_et_al_2014} \\
2013-04-14.1	&	07 42	&	+17 57	&	1.050	&	0.307	&	72.8	&	\cite{Pravec_et_al_2014} \\
2013-04-15.0	&	07 44	&	+18 03	&	1.048	&	0.308	&	73.2	&	\cite{Pravec_et_al_2014} \\
2021-03-18.3	&	08 41	&	+02 08	&	1.080	&	0.120	&	42.9	&	\cite{Warner_2021} \\
2021-03-19.2	&	08 38	&	+02 48	&	1.078	&	0.120	&	44.2	&	\cite{Warner_2021} \\
2021-03-20.3	&	08 35	&	+03 37	&	1.077	&	0.122	&	45.8	&	\cite{Warner_2021} \\
2021-03-21.2	&	08 33	&	+04 15	&	1.076	&	0.123	&	47.1	&	\cite{Warner_2021} \\
2021-03-22.3	&	08 30	&	+05 01	&	1.074	&	0.124	&	48.7	&	\cite{Warner_2021} \\
2021-03-23.3	&	08 28	&	+05 42	&	1.072	&	0.125	&	50.1	&	\cite{Warner_2021} \\
2021-03-24.2	&	08 26	&	+06 19	&	1.071	&	0.126	&	51.4	&	\cite{Warner_2021} \\
2021-03-27.2	&	08 20	&	+08 13	&	1.066	&	0.130	&	55.4	&	\cite{Warner_2021} \\
2021-03-28.2	&	08 19	&	+08 50	&	1.064	&	0.131	&	56.7	&	\cite{Warner_2021} \\
2021-03-29.2	&	08 17	&	+09 25	&	1.062	&	0.133	&	58.0	&	\cite{Warner_2021} \\
2021-03-30.2	&	08 16	&	+09 59	&	1.060	&	0.134	&	59.3	&	\cite{Warner_2021} \\
2021-03-31.2	&	08 14	&	+10 33	&	1.059	&	0.135	&	60.5	&	\cite{Warner_2021} \\

\end{longtable}

\tablefoot{R. A. : Right ascension, Dec. : Declination, $\Delta$ : geocentric distance, $r_{h}$ : heliocentric distance, $\alpha$ : phase angle.}

\clearpage

\section{Light curves}
\label{app:lcs}

   \begin{figure*}[h]
   \centering
    \includegraphics[width=0.8\textwidth, trim=0.25cm 0cm 0cm 0cm, clip]{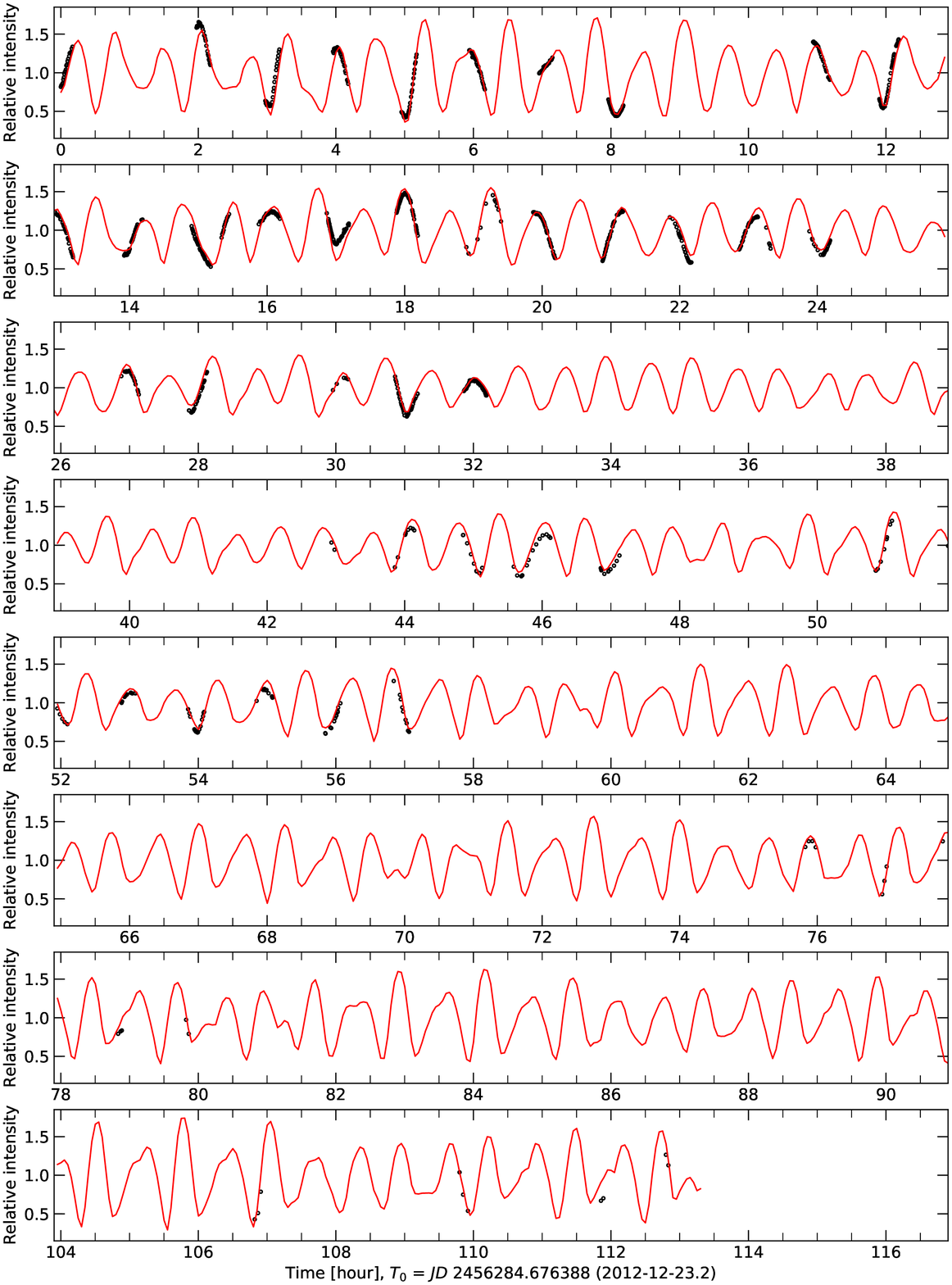}                                             
      \caption{The photometric data in 2012-2013 (black open circle) with the synthetic light curve from best-fit model (red line).}
        \label{fig:lc_appendix_2012}
   \end{figure*}

   \begin{figure*}[t]
   \centering
    \includegraphics[width=0.8\textwidth, trim=0.25cm 0cm 0cm 0cm, clip]{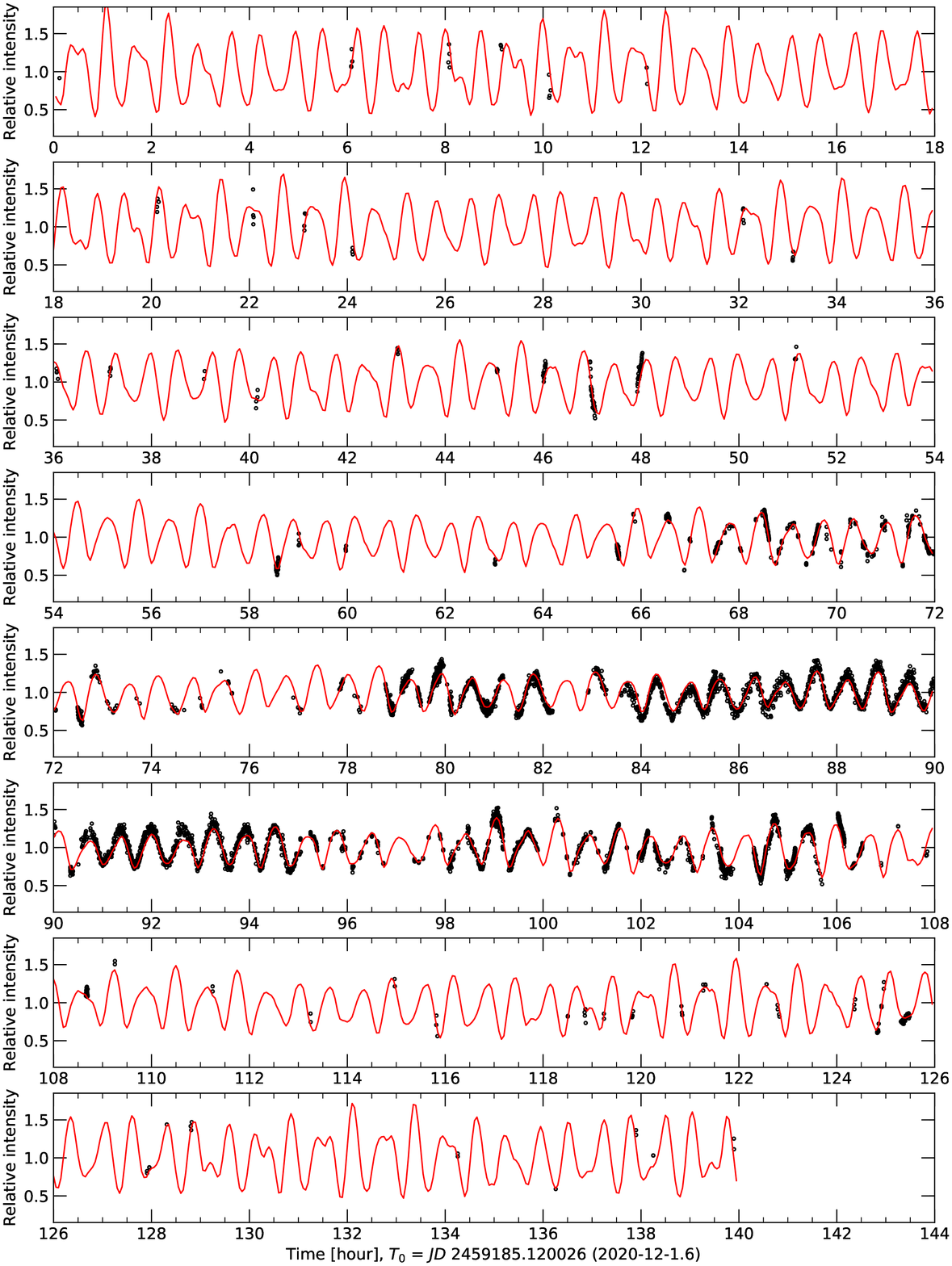}                                             
      \caption{The photometric data in 2020-2021 (black open circle) with the synthetic light curve from best-fit model (red line).}
        \label{fig:lc_appendix_2021}
   \end{figure*}

\end{appendix}

\end{document}